\documentclass[a4paper]{jpconf}
\usepackage{graphicx}
\usepackage{epsfig}
\usepackage{latexsym}
\usepackage{amssymb}
\usepackage{amsbsy}
\usepackage{amsmath}
\usepackage{cite}
\usepackage{epsfig}
\usepackage{graphics,color}
\usepackage{a4}
\newcommand{\as}{\alpha_s}

\newcommand{\be}{\begin{equation}}
\newcommand{\ee}{\end{equation}}

\newcommand{\bea}{\begin{eqnarray}}
\newcommand{\eea}{\end{eqnarray}}
\newcommand{\bean}{\begin{eqnarray*}}
\newcommand{\eean}{\end{eqnarray*}}

\newcommand{\half} {\frac{1}{2}}
\newcommand{\om} {\omega}

\newcommand{\pv}{{\mathbf p}}
\newcommand{\nv}{{\mathbf n}}

\begin{document}
\title {Quark-Gluon Plasma: from lattice simulations to experimental results}
%\subtitle

\author{G.~Aarts, C.~Allton} 

\address{ Department of Physics, College of Science, Swansea University, Swansea, United Kingdom} 

\author{A.~Kelly, J.-I.~Skullerud}

\address{Department of Mathematical Physics, National University of Ireland Maynooth,
Maynooth, County Kildare, Ireland}

\author{S.~Kim}
\address{Department of Physics, Sejong University, Seoul 143-747, Korea} 

\author{T.~Harris, S.~M.~Ryan} 
\address{School of Mathematics, Trinity College, Dublin 2, Ireland} 

\author{M.~P.~Lombardo}
\address{INFN-Laboratori Nazionali di Frascati, I-00044, Frascati (RM) Italy}

\begin{abstract}
Theoretical studies of quarkonia can elucidate some of 
the important properties of the quark--gluon plasma,
the state of matter realised when the temperature exceeds 
$\mathcal{O}(150)$ MeV, currently probed by  heavy-ion collisions experiments at BNL and the LHC.
We report on our results of lattice studies of bottomonia 
for temperatures in the  range $100\,\text{MeV} \lesssim T \lesssim 450\,\text{MeV}$,
introducing and discussing the  methodologies we
have applied. Of particular interest is the analysis of
the spectral functions, where Bayesian methods borrowed and adapted 
from  nuclear and condensed matter physics have proven
very successful. 
\end{abstract}

\section{The plasma of quarks and gluons} 

Heavy ions colliding at ultrarelativistic energies produce a tiny
fireball  of a plasma of quarks and gluons --- the state of matter thought
to have existed slightly after the Big Bang.   
This experimental program started at the SPS in  the 1980s, continued
at RHIC and it is now running at the LHC,  where 
the experiments  ALICE,  ATLAS and CMS
are collecting and analyzing data from the collisions of 
lead nuclei. The most recent runs
reached temperatures of about 500 MeV --- 
approximately $5 \times 10^{12} $ K --- well above the
temperature of the crossover from ordinary matter to the
plasma of quarks and gluons, estimated to be at about 155 MeV \cite{Borsanyi:2010bp,Bazavov:2011nk}.

The analysis of the transition from ordinary matter to the
quark-gluon plasma, 
and the nature of the plasma itself ---
the spectral properties and the residual interactions ---
is a very active field of theoretical research
\cite{FB}. We
are concerned with temperatures well above those where the synthesis
of the lightest nuclei takes place, so strong interactions alone suffice
to describe the system. Hence,  the appropriate microscopic description is the relativistic field 
theory of the strong interactions, Quantum Chromodynamics, QCD.
We also know that  $\alpha_s$, the QCD coupling 
 at the scale of the temperatures of interest, can be as
large as 0.5 and even larger, so that perturbation theory, which
will ultimately be valid for very high temperatures, does not work, at least
quantitatively, in the region which we wish to explore here.
We therefore need a non-perturbative method, and we choose  numerical simulations of
QCD discretized on a lattice.

This note is 
devoted to the presentation of lattice results obtained by the 
FASTSUM collaboration, in particular those appearing in Refs.\ 
\cite{Aarts:2010ek,Aarts:2011sm,Aarts:2012ka,Aarts:2013kaa}.
It updates and expands  previous reviews \cite{Aarts:2012md,Allton:2013uua}.

\subsection {Why bottomonium?}

Why is bottomonium such an interesting probe of the medium? 
We will be concerned with phenomena occurring at, or above, 
the crossover to a chirally 
symmetric, deconfined phase. In general both chiral symmetry restoration and
deconfinement  affect the spectrum of the theory: chiral symmetry will
be seen in the light sector, by the degeneracy of the chiral partners.
The heavier quarks, however, will be   blind to chiral symmetry:
$m_u, m_d \ll m_s \simeq T_c \simeq \Lambda_{QCD} \ll m_c, m_b$.  For 
instance only about 15\% of the strange mass is due to the breaking of
chiral symmetry, and for charm and bottom this contribution is completely negligible --- 
modifications of the spectrum of charmonia and bottomonia come entirely
from the gauge dynamics. 
In very short summary, then, quarkonia  are ideal probes of the
gluodynamics. Since their size is small, however,  the sensitivity to
deconfinement is not immediate: the short-range component
of the potential, which is responsible for their binding, and hence
the fundamental bound states, might well survive in the plasma, while 
the excited states dissolve.  We talk of sequential suppression of quarkonia,
and the goal of our studies is to make quantitative these very qualitative
considerations. 

Charmonia --- which are easier to produce experimentally ---
have been studied since early SPS days. 
We know by now \cite{Rapp:2009my} 
that the experimental results for charmonia are also sensitive
to  cold nuclear matter effects which reduce the primordial
charmonium number significantly --- for instance by about 60\%  at SPS. 
There are also competing temporal scales --- thermalization of the plasma
and formation time of the bound states, as 
well as  the intriguing observation that charmonia
production rates at SPS and RHIC are quite similar. The latter effect 
can be at least in part explained by
 taking into account a regeneration mechanism for charmonia 
(feed-down): the higher
dissociation rate at RHIC is compensated by a richer feed-down. 
All these considerations make the study of charmonium suppression patterns
extremely fascinating but also challenging. 

Bottomonium production, on the other hand,
is much less prone to regeneration effects, and as such  
is a more promising observable for the 
spectral analysis of the quark--gluon plasma\cite{Rapp:2009my}. 
Since  bottomonium production
requires the larger energies available at the LHC, 
such data have become available only very recently. In the last few years, 
results showing sequential Upsilon suppression
in PbPb collision at LHC energies have appeared 
\cite{Chatrchyan:2011pe,Chatrchyan:2012lxa},
and we will comment on those at the end.

\section{Lattice QCD, relativistic and non--relativistic}

Lattice calculations are performed in a discretized Euclidean 
space-time, which introduces technical scales: the lattice spacing $a$
and the lattice spatial size, $L$. Each characteristic 
physical scale 
{\it l}  should obviously fulfil 
the constraint $ a \ll l \ll L$. Accommodating quarks
with vastly different masses then  poses a computational challenge.

Our strategy is  to treat as accurately as possible
the light quarks: 
in all our studies gauge configurations 
with dynamical light Wilson-type quark flavours are produced on 
highly anisotropic lattices.
Details of the lattice action and parameters can be found in 
Refs.~\cite{Morrin:2006tf,Oktay:2010tf}. 
For the $b$ quarks 
we use non-relativistic QCD (NRQCD). 
In our work we use a  mean-field improved action with tree-level 
coefficients, which includes
terms up to and including ${\cal O}(v^4)$,
where $v$ is the typical velocity of a bottom quark in 
bottomonium,$v^2_b \simeq 0.1$.
There is no (rest) mass term in the NRQCD action so one  can dispense with
the demanding constraint $ a \ll 1/m_b$. In general,
NRQCD relies on the separation of scales between  the bottom  quark
and any other physical scale of the theory: 
in our work we study temperatures up to $2 T_c \simeq 400$ MeV, hence
$m_b \gg T$ and the application of NRQCD  is  fully justified.

\section{ Correlators in the plasma}

In our studies 
we considered  the  $S$ wave states $\Upsilon$ and $\eta_b$ and 
the  $P$ wave  states $\chi_{b0}$, $\chi_{b1}$, $\chi_{b2}$ and
$h_b$. We found that the correlators in the different $P$ wave channels 
behave in a very similar way, 
hence from now on for the $P$ wave states 
we present results for the $\chi_{b1}$ channel only.
The gauge configurations are generated with
a dynamical first
generation of quarks. More recently the strange quark
has been added as well\cite{Aarts:2014cda}: the general features of the results
are unchanged, and a systematic, detailed analysis  
of the effect of a dynamical strange quark is in progress.

\begin{figure}[t]
{\includegraphics[height=5.0cm, width=4.2cm]{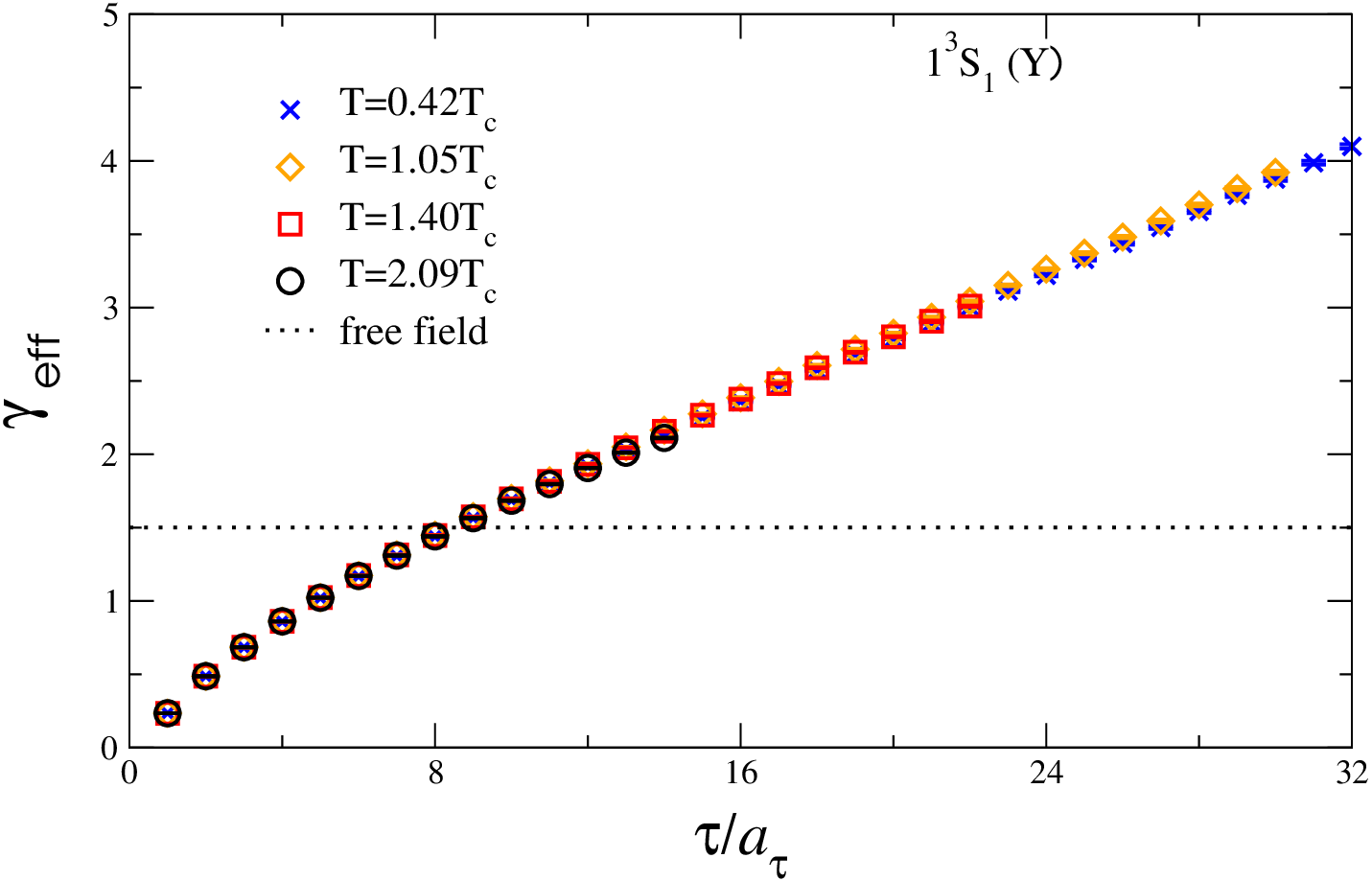}}
{\includegraphics[height=5.0cm,width=4.2cm]{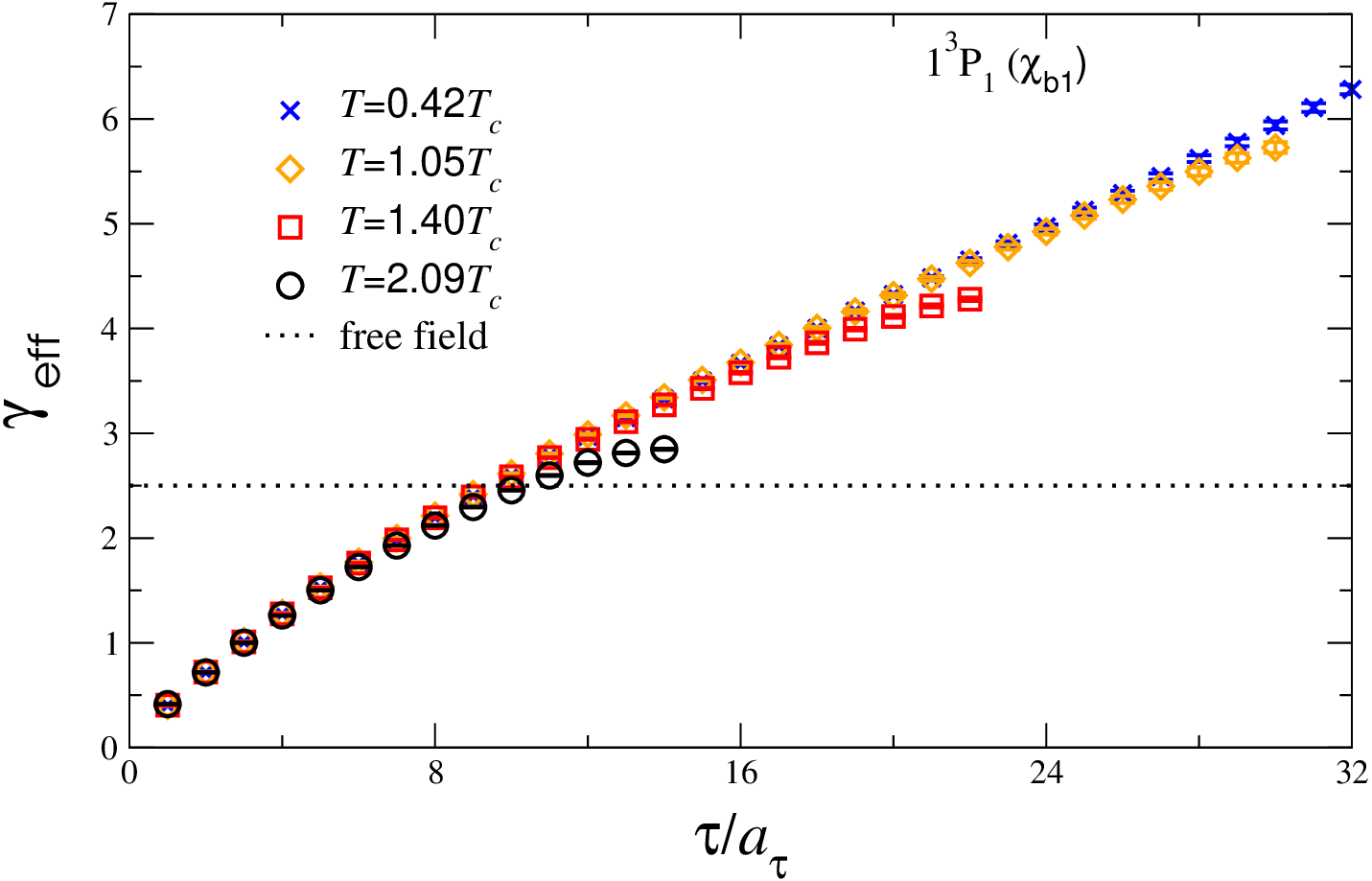}}
{\includegraphics[height=5.0cm,width=4.2cm]{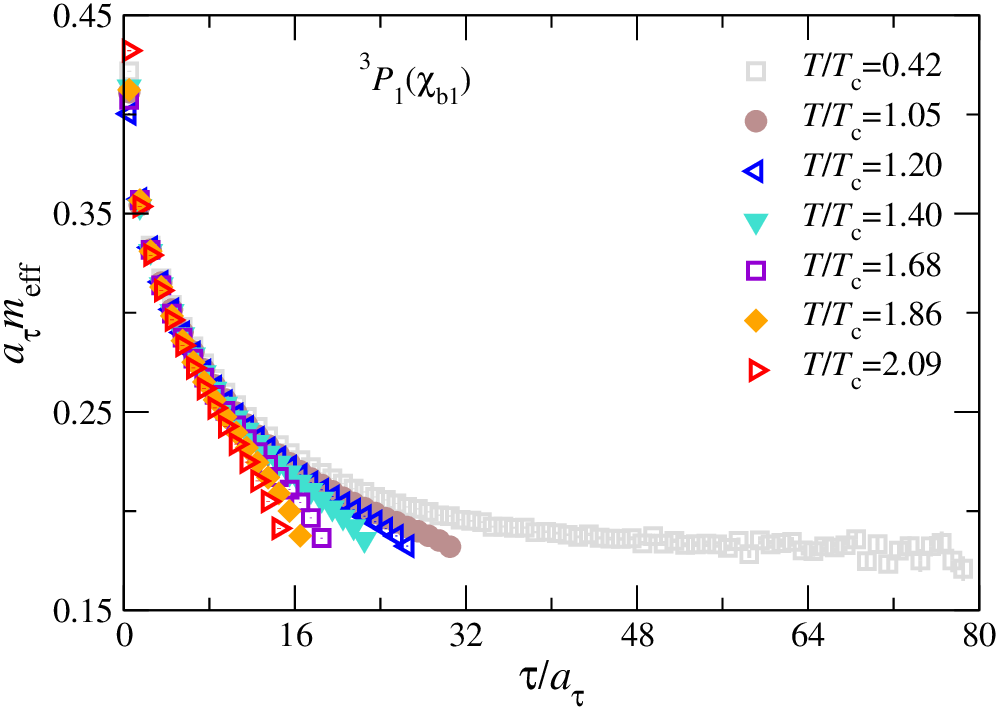}}
\caption{ Effective exponents $\gamma_{\rm eff}(\tau)$ for the
  $\Upsilon$ (left) and $\chi_{b1}$ (center), as a function of
  Euclidean time for various temperatures. The dotted line indicates
  the non-interacting  result in the continuum, which is approached by
  the $\chi_{b1}$ results at the higher temperature $T= 2.1 T_c$ \cite{Aarts:2010ek}. 
  Effective mass for the $\chi_{b1}$ (right),  as a function of Euclidean time for various
  temperatures \cite{Aarts:2011sm}. } 
\label{fig:power}
\end{figure}

There is a further important simplification in NRQCD: 
to compute propagators we only need to solve an initial value problem,
in contrast to the relativistic case where the same computation 
requires the inversion of a large sparse matrix. 
In the hadronic  phase we expect that the large Euclidean time behaviour
of the propagators is dominated by an exponential decay. 
In the QGP phase, one can use as guidance the behaviour found for free quarks in NRQCD
\cite{Burnier:2007qm,Aarts:2010ek,Aarts:2011sm}, which yields the spectral functions
\begin{align}
    \rho_{\textrm{free}}(\omega)\propto(\omega-\omega_0)^{\alpha}\,\Theta(\omega-\omega_0),
        \quad\textrm{where}\quad
        \alpha=
        \begin{cases}
            1/2, &\textrm{S~wave}.\\
            3/2, &\textrm{P~wave}.
        \end{cases}
    \label{eq:rhofree}
\end{align}
Here we supplemented \cite{Aarts:2014cda} the free results by a threshold, $\omega_0$, to account for the additive shift in the quarkonium energies which describes the residual interactions in the thermal
medium.  The correlation functions then have the following behaviour
\begin{align}
    G_{\mathrm{free}}(\tau)\propto \int d\omega\, e^{-\omega \tau}\rho(\omega)
     \propto \frac{e^{-\omega_0\tau}}{\tau^{\alpha+1}}.
    \label{eq:powerlaw}
\end{align}
To visualize the temperature dependence and at the same time 
monitor the approach to a quasi-free behaviour
we construct effective power plots \cite{Aarts:2010ek}, using the definition 
\begin{equation}
\gamma_{\rm eff}(\tau) = -\tau\frac{G'(\tau)}{G(\tau)}
 = -\tau\frac{G(\tau+a_\tau) - G(\tau -a_\tau)}{2a_\tau G(\tau)},
\end{equation}
where the prime denotes the (discretized) derivative.  For a pure power
law decay this yields a constant,
 $\gamma_{\rm  eff}(\tau)=\alpha + 1$. Taking into account the threshold
$\omega_0$, one finds $\gamma_{\rm  eff}(\tau)=\alpha + 1 + \omega_0\tau$.
  The results are shown in
 Fig.~\ref{fig:power}.    We see that the $\Upsilon$ displays a very
 mild temperature  dependence, while for the $\chi_{b1}$ the 
asymptotic behaviour of the effective
 power tends to flatten  out with increasing temperature: this
indicates that $\omega_0$ --- the slope ---  
decreases with temperature.  In the same panel  also shown are the
 effective exponents in the continuum non-interacting limit
($\omega_0 =0$).  In the
 case of the $\chi_{b1}$, we observe that the  effective exponent ---
which can be read off the intercept of the asympotic straight line
with the $\tau=0$ axis ---  seems to
approach the non-interacting result. 
 Note that, even for $T \to \infty$ where $\omega_0 = 0$,   the effective
exponents on the lattice only become a straight line, $\gamma_{\rm  eff}(\tau)\to \alpha + 1$, at large enough $\tau$ due to the presence of lattice discretisation effects at smaller temporal separations.

A complementary description  is offered by the
 effective masses,
\begin{equation}
a_\tau m_{\rm eff}(\tau) = - \log[G(\tau)/G(\tau-a_\tau)],
\end{equation}
which are shown in the rightmost panel of Fig.\ \ref{fig:power}.  When
the correlator takes the form of a sum of exponentials, the ground
state will show up as a plateau at large Euclidean times, provided
that it is well separated from the excited states. This is indeed the
case at the lowest temperature and leads to the zero-temperature
spectrum discussed in Refs.\  \cite{Aarts:2010ek,Aarts:2011sm}. Above
$T_c$, we observe that the effective masses no longer follow the trend
given by the correlator below $T_c$, but instead bend away from the
low-temperature data.
The results shown in Fig.~\ref{fig:power} imply  that the spectrum of
the $\chi_{b1}$ has changed drastically.  If isolated bound states
persist, the ground state has to  be much lighter and excited states
cannot be well separated.  A more natural explanation is  that there
is no exponential decay  and bound states have melted,  immediately
above $T_c$.  This interpretation is supported by the spectral
function  analysis presented next.

\section{Spectral functions}

Spectral functions  play an important role in understanding how
elementary excitations are modified in a thermal medium, from many-body 
physics --- see the talk by Giuseppina Orlandini at this meeting
\cite{GO} --- to QCD, which is discussed here.

\begin{figure}[!t]
\begin{center}
\includegraphics*[width=12cm]{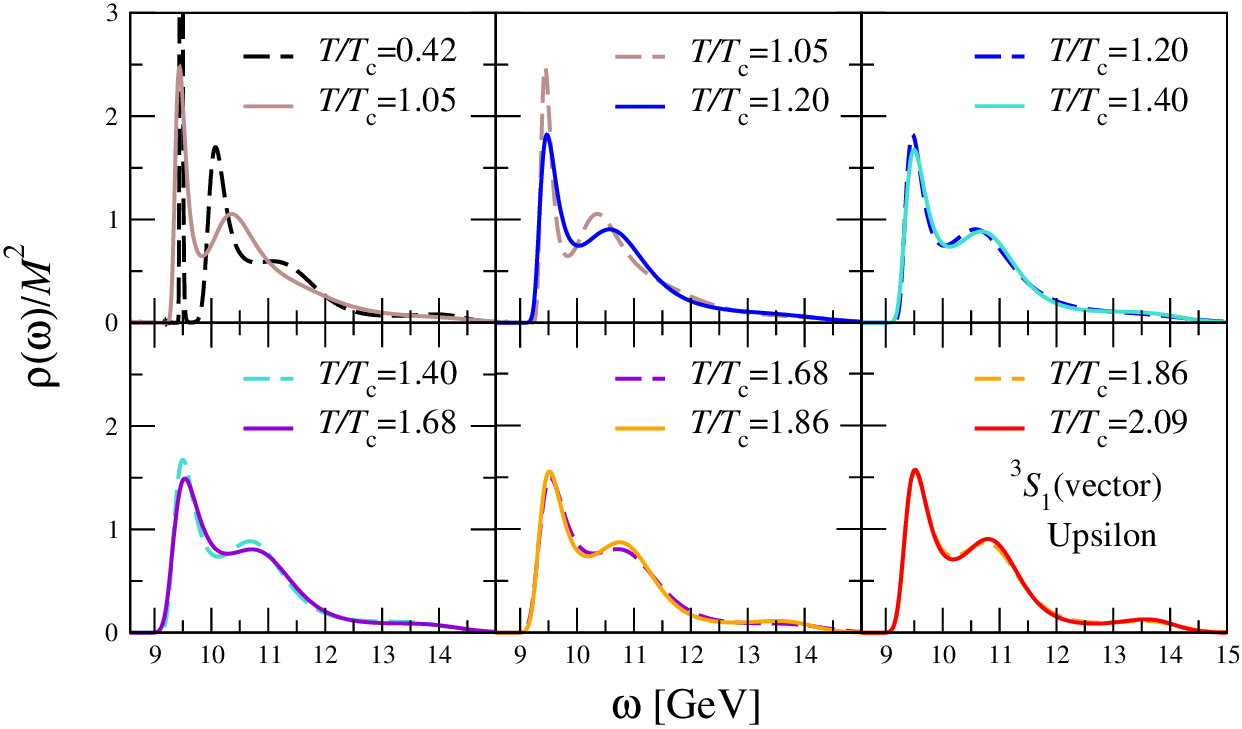}
\includegraphics*[width=12cm]{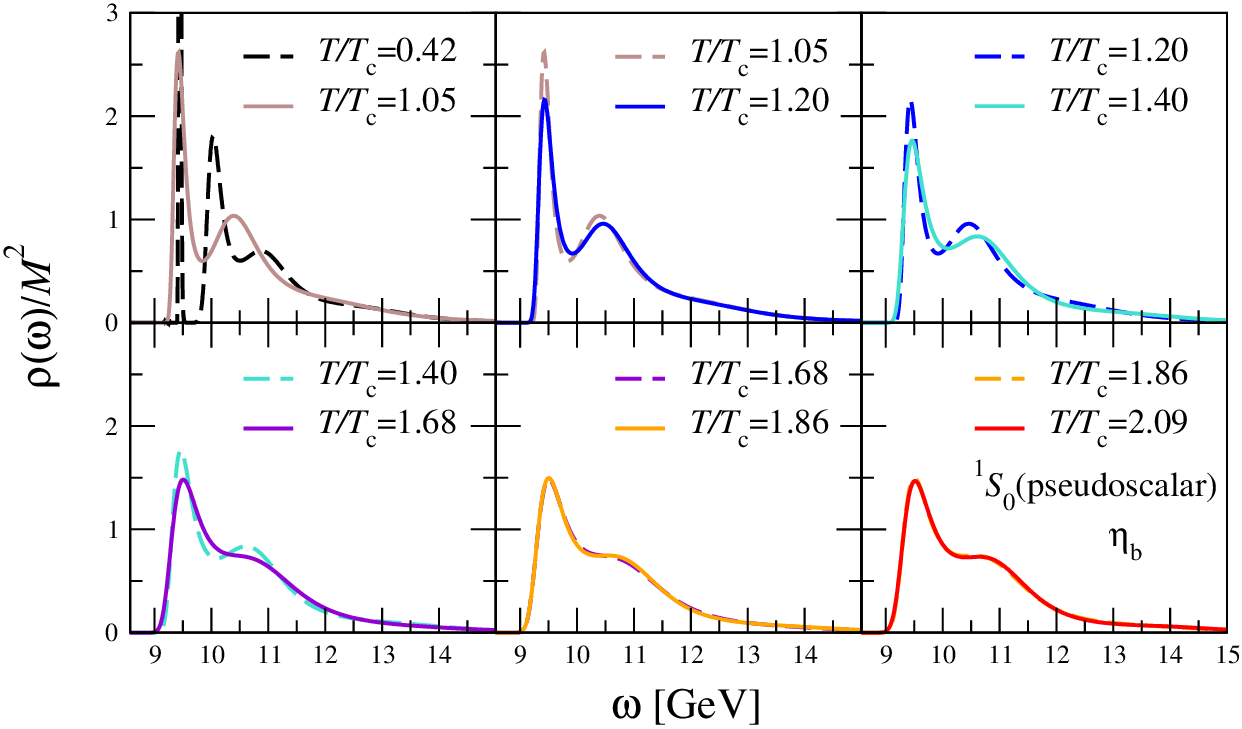} 
\end{center}
\caption{ Spectral functions $\rho(\omega)$, normalised with the heavy
  quark mass, in the vector ($\Upsilon$) channel  (upper panel)  and
  in the pseudoscalar ($\eta_b$) channel (lower panel) for all
  temperature available. The subpanels are ordered from cold (top
  left) to hot (bottom right). Every subpanel contains two adjacent
  temperatures to facilitate the comparison \cite{Aarts:2011sm}.  }
\label{fig:rho_all_upsilon}
\end{figure}

In general, the spectral decomposition of a (relativistic) zero-momentum  
Euclidean propagator $G(\tau)$ at finite temperature $T$  is given by
\begin{equation}
\label{eq:K}
G(\tau) = \int_{-\infty}^\infty \frac{d\omega}{2\pi}\,  K(\tau,\om)
\rho(\omega),
\end{equation}
where $\rho(\omega)$ is the spectral function and  the kernel $K$ is
given by
\begin{equation}
K(\tau,\om) =
\frac{\cosh\left[\omega(\tau-1/2T)\right]}{\sinh\left(\omega/2T\right)}.
\end{equation}
In NRQCD the kinematical temperature dependence is always absent. This
can be seen in a number of ways. Following
Ref.\  \cite{Burnier:2007qm}, we write $\om=2M+\om'$ and drop terms
that  are exponentially suppressed when $M\gg T$. The spectral
relation (\ref{eq:K}) then reduces to its zero-temperature limit,
\begin{equation}
G(\tau) = \int_{-2M}^\infty\frac{d\omega'}{2\pi}\, \exp(-\om'\tau)\rho(\omega'),
\end{equation}
even at nonzero temperature.  Since the interesting physics takes place around the two-quark threshold, $\om\sim 2M$, the region of interest is around $\omega'\sim 0$ and the lower limit becomes irrelevant.
 In summary, in the heavy-quark limit the
spectral relation simplifies considerably, and temperature effects
seen in the correlators are thus only due to changes in the
light-quark--gluon system.

Despite these simplifications, the calculation of the  NRQCD spectral
functions using Euclidean propagators as an input remains a difficult,
ill-defined problem.  We will tackle it by   using the Maximum Entropy
Method (MEM) \cite{Asakawa:2000tr}, which has proven successful in a
variety of applications.  We have carefully studied the systematics,
including the dependence on the set of lattice data points in time,
and on the  default model  $m(\om)$ which enters in the
parametrisation of the spectral function, 
\begin{equation}
\rho(\om) = m(\om) \exp \sum_k c_ku_k(\om),
\end{equation}
where $u_k(\om)$ are basis functions fixed by the kernel $K(\tau,\om)$
and the number of time slices, while  the coefficients $c_k$ are to be
determined by the MEM analysis \cite{Asakawa:2000tr}. We find that the
results are insensitive to the choice of default model, provided that
it is a smooth function of $\om$.  It remains of interest, and it is
an open avenue of research, to experiment with alternative
prescriptions \cite{Rothkopf:2011ef,Burnier:2013nla}, also on
ensembles generated with different lattice actions \cite{seyong}.

The results for the spectral functions for  the $S$ wave states,
$\Upsilon$  and  $\eta_b$ \cite{Aarts:2011sm} can be seen in
Fig.\ \ref{fig:rho_all_upsilon}, which shows that as the temperature
is increased the ground state peaks of both states remain visible.
One caveat applies also here: the apparent width at zero temperature
is most likely due to a lattice/MEM artefact, and this calls for an
analysis of the discretization effects on the spectral function, which
is one of our ongoing projects \cite{Harris:2013vfa}.  The peaks
associated with the  excited
states become suppressed at higher temperature and are no longer
discernible quickly above $T_c$.

\begin{figure}[t]
\begin{center}
\epsfig{figure=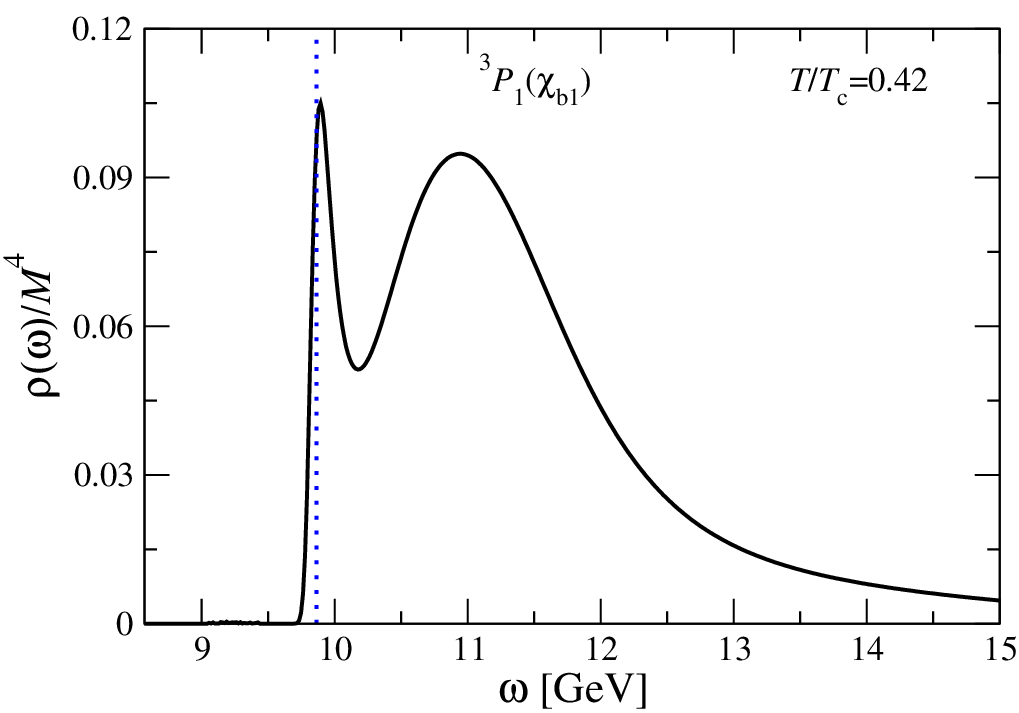,height=5.0cm,width=0.32\textwidth}
\epsfig{figure=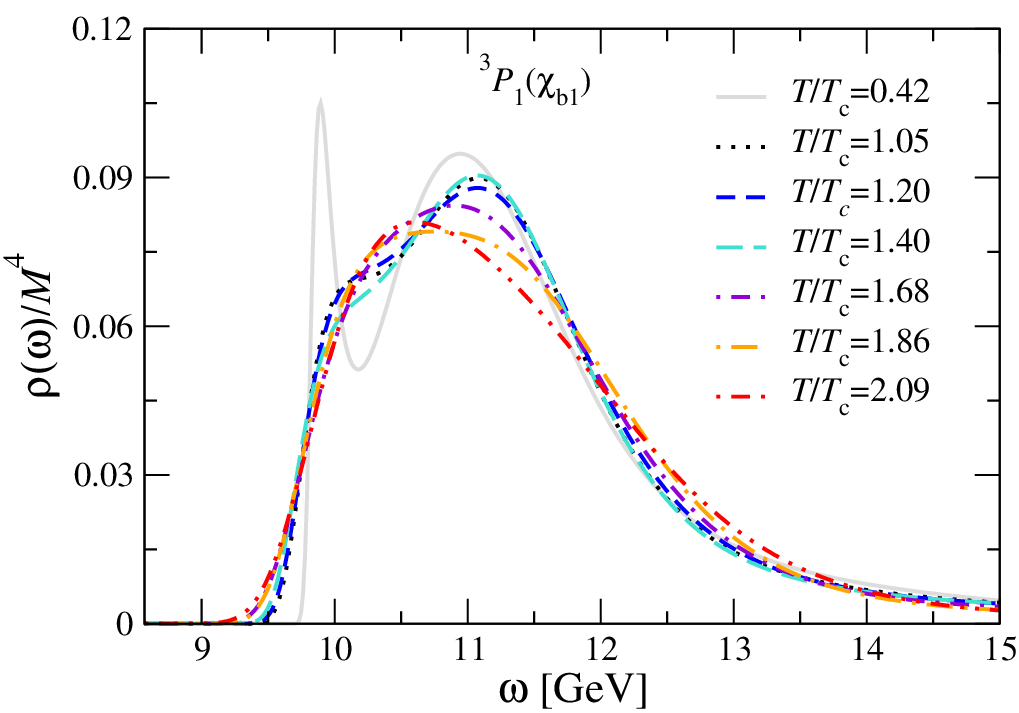,height=5.0cm,width=0.32\textwidth}
\epsfig{figure=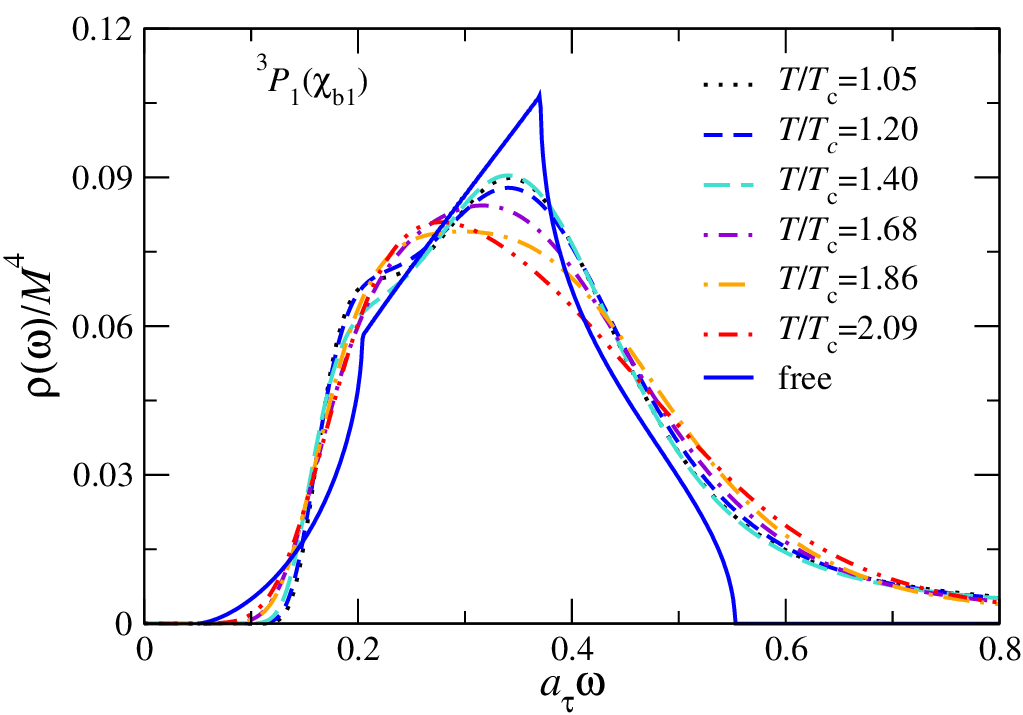,height=5.0cm,width=0.32\textwidth}
\end{center}
\caption{Spectral function in the $\chi_{b1}$ channel at the lowest
  temperature (left) and at all temperatures (center). The dotted line
  on the left indicates the position of the ground state  obtained
  with a standard exponential fit. Comparison with the free lattice
  spectral function above $T_c$ (right) \cite{Aarts:2013kaa}.}
\label{fig:2}
\end{figure}

Turning now to the $P$ wave states,  the result at the lowest
temperature is given in Fig.\ \ref{fig:2} (left). The dotted vertical
line indicates the mass of the lowest-energy state obtained with an
exponential fit. We see from this that the narrow peak in the spectral
function corresponds to the ground state.  The second, wider structure
is presumably a combination of excited states and lattice artefacts,
see below. We note that we have not been able to extract the mass of
the first excited state with an exponential fitting procedure.  The
spectral functions for all temperatures are shown in Fig.\ \ref{fig:2}
(center). We  find no evidence of a ground state peak for any of the
temperatures above $T_c$. This is consistent with the interpretation
of the correlator study presented above and supports the conclusion
that the $P$ wave bound states melt in the QGP.

In order to interpret the remaining structure, we compare it with the
spectral function computed on the lattice, in the absence of
interactions \cite{Aarts:2011sm}.   In Fig.\ \ref{fig:2},
rightmost panel,  we show the free lattice spectral functions,
together with the spectral functions above $T_c$. In that plot we
have adjusted the threshold to match our lattice results, and once
this is done the lattice spectral function and the free one almost
coincide.   This lends further support to the conclusion drawn in
Ref.\ \cite{Aarts:2010ek} from an analysis  of the correlators: 
the system in the $P$ wave channels is approaching a system of
noninteracting quarks, and the residual interactions can be described
by a shift  in the quarkonium energies, which does not affect the
shape of spectral functions.

\section{Comparison with analytic studies and momentum dependence}

In this section we restrict our analysis to the $S$ waves.  From the
computed spectral functions we can determine the mass (from the peak
position) and (an upper bound on) the width of the ground state at
each  temperature. 

\begin{figure}[!t]
\begin{center}
\includegraphics*[height=4cm]{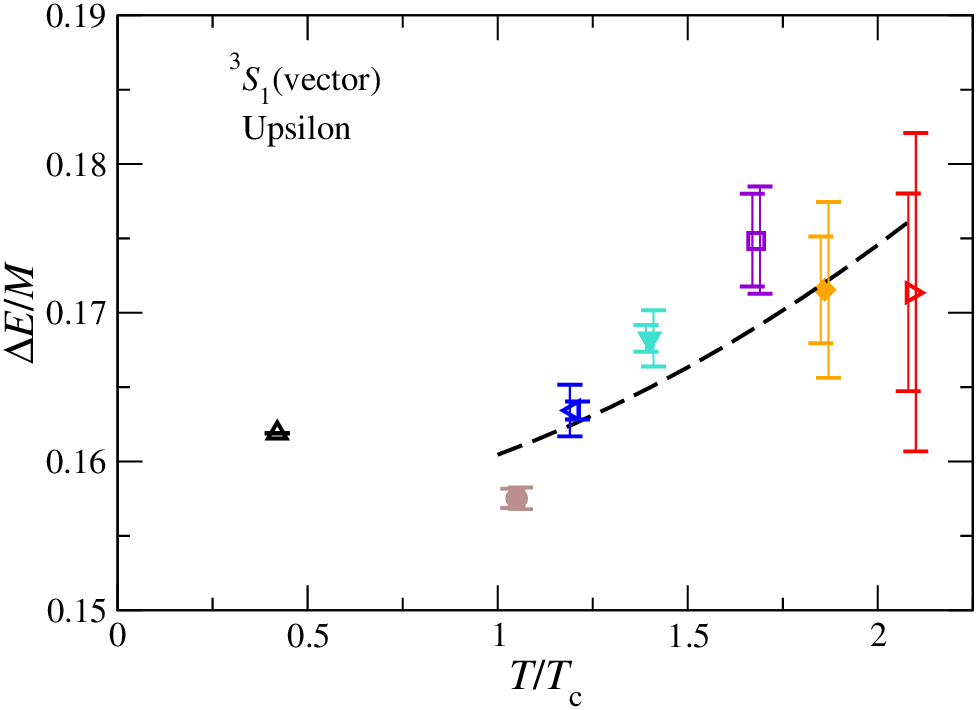}
\includegraphics*[height=4cm]{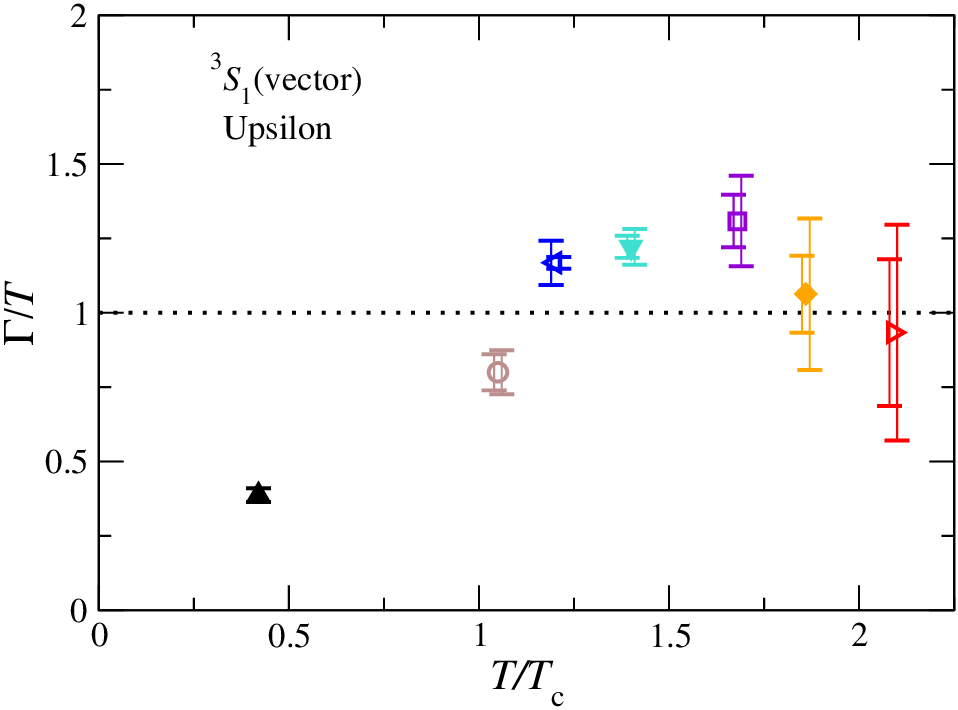}
\end{center} 
\caption{ Position of the ground state peak $\Delta E$, normalised by
  the heavy quark mass (left),  and the upper limit on the width of
  the ground state peak, normalised by the temperature (right),  as a
  function of $T/T_c$ in the vector ($\Upsilon$) channel.  Similar
  results have been obtained  for the pseudoscalar ($\eta_b$) channel
  \cite{Aarts:2011sm}.  }
\label{fig:mass}
\end{figure}

In Fig.\ \ref{fig:mass} we show the temperature dependence of the mass
shift $\Delta E$, normalised by the heavy quark mass and the
temperature dependence of the width, normalised by the temperature.

We now  contrast our results with analytic predictions derived
assuming a weakly coupled plasma \cite{Burnier:2013nla,
Laine:2006ns,
Laine:2007gj,
Brambilla:2008cx,
Brambilla:2010vq,
Brambilla:2011sg,
Escobedo:2011ie}.  According to
Ref.\ \cite{Brambilla:2010vq}, the thermal contribution to the width
is given, at leading order in the weak coupling and large mass
expansion, by
\begin{equation}
\frac{\Gamma}{T} =  \frac{1156}{81}\as^3 \simeq 14.27\as^3,
\end{equation}
i.e. the width increases linearly with the temperature.  If we take
as an estimate from our results that  $\Gamma/T\sim 1$, we find that
this corresponds to $\as\sim 0.4$, which is a reasonable result.  (It
would be of interest to compute $\as$ directly on our configurations.)
In the same spirit  the thermal mass shift is given in
Ref.~\cite{Brambilla:2010vq} by
\begin{equation} 
\delta E_{\rm thermal} =  \frac{17\pi}{9}\as\frac{T^2}{M} \simeq
5.93\as\frac{T^2}{M}. \label{eq:deltaE-pert}
\end{equation}
In these simulations we have $T_c\sim 220$ MeV, $M\sim 5$ GeV.  Taking
these values together with $\as\sim 0.4$ as determined above,
Eq.\ \eqref{eq:deltaE-pert} becomes
\begin{equation}
\frac{\delta E_{\rm thermal}}{M} =5.93\as
\left(\frac{T_c}{M}\right)^2\left(\frac{T}{T_c}\right)^2 \sim
0.0046\left(\frac{T}{T_c}\right)^2.
\end{equation}
In order to contrast our results with this analytical prediction, we
have compared the temperature dependence of the peak positions to the
simple expression
\begin{equation}
\frac{\Delta E}{M} =c+ 0.0046\left(\frac{T}{T_c}\right)^2, 
\end{equation}
where $c$ is a free parameter. This is shown by the dashed line in
Fig.\ \ref{fig:mass} (left panel). The numerical results and the
analytic ones are not inconsistent, within the large errors.

\begin{table}[t]
\begin{center}
\begin{tabular}{| l | ccccccc | }
\hline $\nv$ 			& (1,0,0)	& (1,1,0)   & (1,1,1)
&  (2,0,0)	& (2,1,0)	& (2,1,1)	& (2,2,0)
\\ $|\pv|$ (GeV)		& 0.634    & 0.900 	& 1.10 	& 1.23
& 1.38 	& 1.52 	& 1.73	\\ $v$ [$\Upsilon(^3S_1)$]	& 0.0670
& 0.0951	& 0.116	& 0.130	& 0.146	& 0.161 	& 0.183
\\ $v$ [$\eta_b(^1S_0)$]	& 0.0672 	& 0.0954 	&
0.117 	& 0.130 	& 0.146 	& 0.161 	& 0.183
\\ \hline
\end{tabular}
\vspace*{0.2cm}
\caption{Nonzero momenta \cite{Aarts:2012ka}. Also indicated are the
  corresponding velocities $v=|\pv|/M_S$ of the ground states in the
  vector ($\Upsilon$) and pseudoscalar ($\eta_b$) channels, using the
  ground state masses determined previously  \cite{Aarts:2010ek},
  $M_{\Upsilon}=9.460$ GeV and  $M_{\eta_b}=9.438$ GeV.  }  
\label{tab:mom}
\end{center}
\end{table}

The analysis from effective theories predicts significant momentum
effects at large momenta. Moreover current CMS results have been
obtained at large momenta. There is therefore both phenomenological and
experimental motivation to extend these studies to non-zero
momenta\cite{Aarts:2012ka}.  
The momenta and velocities that are accessible on the lattice are
constrained by the discretization and the spatial lattice spacing. The
lattice dispersion relation  reads
\begin{equation}
a_s^2\pv^2 = 4\sum_{i=1}^3 \sin^2\frac{p_i}{2}, \quad\quad p_i =
\frac{2\pi n_i}{N_s}, \quad\quad -\frac{N_s}{2} < n_i \leq
\frac{N_s}{2}.
\end{equation}
To avoid lattice artefacts, only momenta with $n_i<N_s/4$ are used: we
consider the combinations (and permutations thereof) given in Table
\ref{tab:mom}.  The largest momentum, using $\nv = (2,2,0)$, is $|\pv|
\simeq 1.73$ GeV, corresponding to $v = |\pv|/M_S \simeq
0.2$. Therefore, the range of velocities we consider is
non-relativistic.

\begin{figure}[t]
\begin{center}
\epsfig{figure=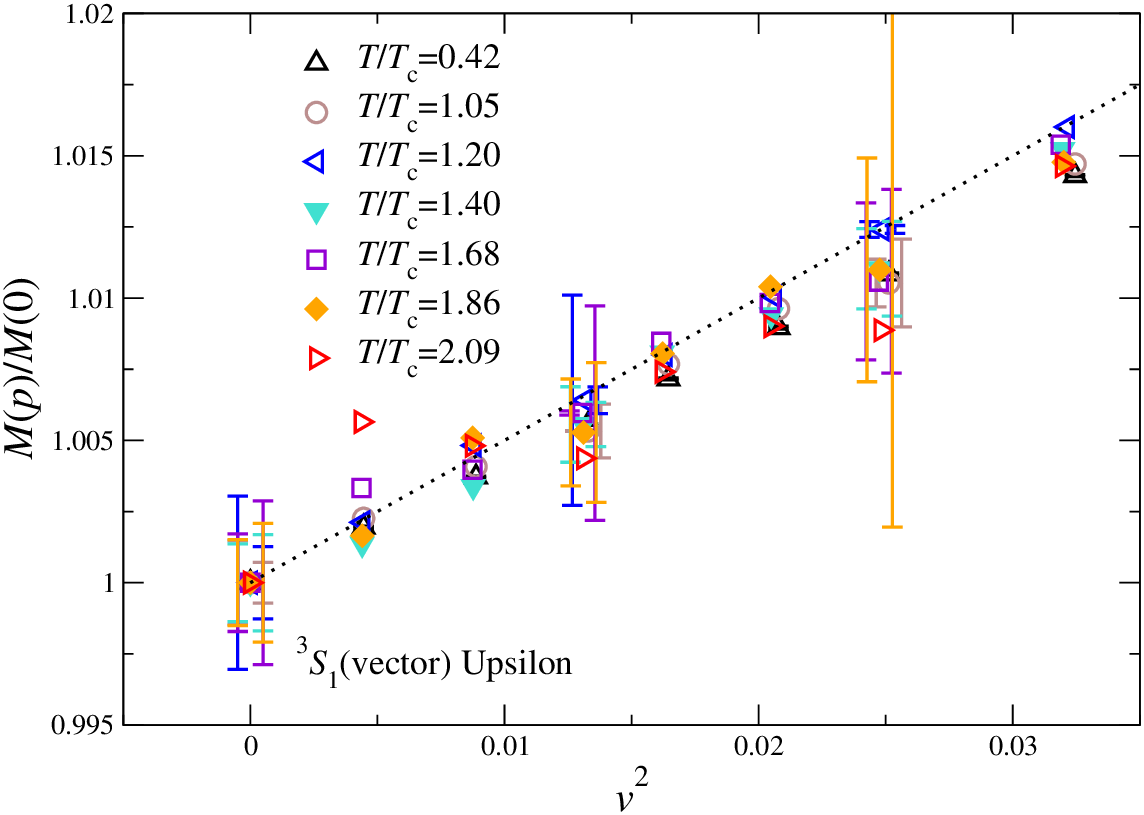,width=0.4\textwidth}
\epsfig{figure=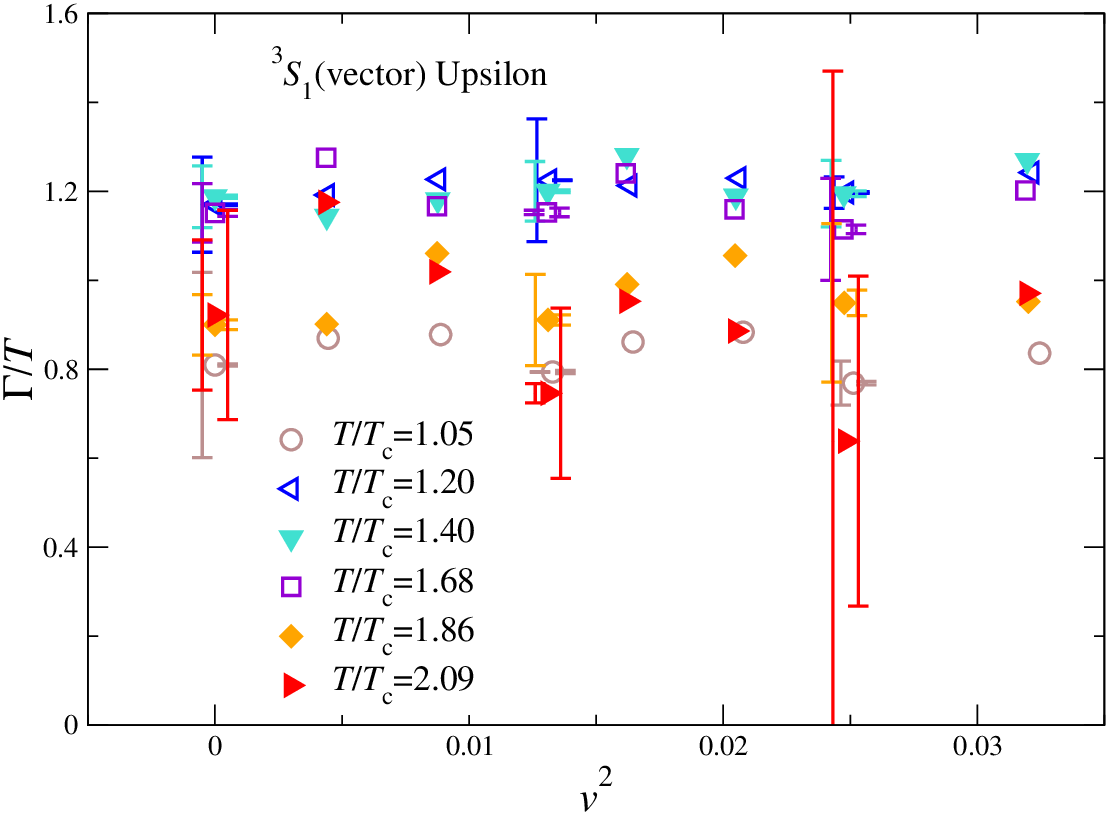,width=0.4\textwidth}
\end{center}
\caption{ Position of the ground state peak $M(\pv)/M(0)$ (left) and
  the upper limit on the width of the ground state peak, normalized by
  the temperature,  $\Gamma/T$ (right), as a function of the velocity
  squared ($v^2$) in the vector ($\Upsilon$) channel %(above).
  Analogous results have been obtained for the  the pseudoscalar
  ($\eta_b$) channel.  The dotted line in the left figure represents
  $M(\pv)/M(0)=1+\half v^2$ \cite{Aarts:2012ka}.  }
\label{fig:mass-width}
\end{figure}

As in the zero-momentum case, we extract masses and widths from our
correlators using MEM, and we display them in Fig.\ 5.    We observe
that the peak position increases linearly with $v^2$, as
expected. Assuming the lowest-order, non-relativistic expression
$M(\pv)=M(0) + \pv^2/2M(0)$, one finds
\begin{equation}
\frac{M(\pv)}{M(0)} = 1+ \frac{\pv^2}{2M^2(0)} = 1+\half v^2,
\end{equation}
which is indicated with the dotted lines in the left figures.

The dependence on the velocity  can be compared with EFT
predictions. In Ref.\ \cite{Escobedo:2011ie} a study of the velocity
dependence was carried out  in the context of QED, working in the rest
frame of the bound state (i.e.\ the heat bath is moving). In order to
compare with our setup, we consider the case in which the temperature
is low enough for bound states to be present and where the velocities
are non-relativistic. In that case, one finds \cite{Escobedo:2011ie},
in the rest frame of the bound state and at leading order in the EFT
expansion, 
\begin{equation}
\frac{\Gamma_v}{\Gamma_0}  =
\frac{\sqrt{1-v^2}}{2v}\log\left(\frac{1+v}{1-v}\right),
\end{equation}
where $\Gamma_0$ is the width at  rest.  Interpreting the width as an
inverse lifetime, one can express this result  in the rest frame of
the heat bath by dividing with the Lorentz factor $\gamma =
1/\sqrt{1-v^2}$. An expansion for non-relativistic velocities then
yields
\begin{equation}
\frac{\Gamma_v}{\Gamma_0} =   1 -\frac{2v^2}{3} +  {\cal
  O}\left(v^4\right).
\end{equation}
If we apply this result  to our study of bottomonium,  we find that
the effect of the nonzero velocity shows up as a correction at the
percent level (recall that $v^2\lesssim 0.04$), which is beyond our
level of precision but consistent with the observed $v$ independence
within errors.  Similarly, additional thermal effects in the
dispersion relation are currently beyond our level of precision.  In
summary, the observations in our low-momentum range are consistent with
Ref.\ \cite{Escobedo:2011ie}, and in  order to observe the predicted
non-trivial momentum dependence we need to explore  larger momenta.

\section{From lattice to experiments}
\label{sec:sum}

We have presented our results for bottomonium in the quark-gluon
plasma, for temperatures up to  $2.1T_c$, at the threshold of the
region currently  explored by LHC heavy-ion experiments.  Our
analysis uses  full relativistic dynamics for the light quarks, and a
non-relativistic approach for the bottom quarks. The results are
amenable to a successful comparison with effective models which we
hope to further pursue in the future. 

Our  results  demonstrate a pattern of suppression of bottomonia which
apparently compares  well with recent CMS  results \cite{Chatrchyan:2011pe,Chatrchyan:2012lxa},
once we take into account that the temperatures reached in the collisions
studied by CMS are similar to the ones of our lattices.
However the systems under investigation are  vastly different: 
Quantum Chromodynamics in thermal equilibrium in one case, and an expanding fireball, with an extremely complex
experimental setup, in the other. For instance, our studies include processes 
between a $b$ quark and thermal light quarks (and gluons), but do not include 
the thermal scattering of $b$ quarks. Unravelling the details and the
limitations of this comparison is a subject of active research,
certainly beyond the scope of this note. 

\ack

We thank Don Sinclair and Bugra Oktay for collaboration.
This work was partly supported by the European Community under the FP7 programme HadronPhysics3.
We acknowledge the support and infrastructure provided by the Trinity Centre for High Performance
Computing and the IITAC project funded by the HEA under the Program for Research in Third Level Institutes (PRTLI) co-funded by the Irish Government and the European Union.  
The work of CA and GA is carried out as part of the UKQCD collaboration and the STFC funded DiRAC Facility.
GA and CA are supported by STFC.  
GA is supported by the Royal Society, the Wolfson Foundation and the Leverhulme Trust. 
SK is grateful to STFC for a Visiting Researcher Grant and supported by the National Research
Foundation of Korea grant funded by the Korea government (MEST) No.\
2012R1A1A2A04668255.  
SR is supported by the Research Executive Agency (REA) of the European Union under Grant Agreement number PITN-GA-2009-238353 (ITN STRONGnet) and the Science Foundation Ireland, grant no.\ 11-RFP.1-PHY-3201.

\end{document}